# A neuronal dynamics study on a neuromorphic chip

Wenyuan Li, Igor V. Ovchinnikov, Honglin Chen, Zhe Wang, Albert Lee, Hochul Lee, Carlos Cepeda, Robert N. Schwartz, Karlheinz Meier, and Kang L. Wang

Neuronal firing activities have attracted a lot of attention since a large population of spatiotemporal patterns in the brain is the basis for adaptive behavior and can also reveal the signs for various neurological disorders including Alzheimer's, schizophrenia, epilepsy and others. Here, we study the dynamics of a simple neuronal network using different sets of settings on a neuromorphic chip. We observed three different types of collective neuronal firing activities, which agree with the clinical data taken from the brain. We constructed a brain phase diagram and showed that within the weak noise region, the brain is operating in an expected noise-induced phase (N-phase) rather than at a so-called self-organized critical boundary. The significance of this study is twofold: first, the deviation of neuronal activities from the normal brain could be symptomatic of diseases of the central nervous system, thus paving the way for new diagnostics and treatments; second, the normal brain states in the N-phase are optimal for computation and information processing. The latter may provide a way to establish powerful new computing paradigm using collective behavior of networks of spiking neurons.

## 1. Introduction

One of the most important hypotheses in neural dynamics in the past decades is "the self-organized criticality". Brain, as a complex system with a tremendous large number of elements, brings a lot of attention for its collective dynamic process and behaviors of firing activities [1]. "The criticality hypothesis" states that the brain is poised in the critical boundary between two different dynamic states [2, 3, 4, 5, 6, 7]. Experimental results and theoretical models for this criticality have been found to be rather ubiquitous across multiple species including rats [8], adult cats [9], non-human primates [10,11], and spanning across different measurements such as EEG [12], MEG [13,14,15], LFP [16] and fMRI [17,18]. Furthermore, the different characteristics of neural dynamics behavior can be used as markers of mental diseases. For instance, fMRI studies have been used to evaluate various pathological conditions such as Alzheimer's [19], schizophrenia [20], and epilepsy [21,22], showing a characteristic deviation from the power law. Likewise, EEG data of epileptic dynamics show super-critical state behavior and deviates from the power-law statistics of a normal brain dynamics [23,24].

Though such criticality in neurodynamical behavior has been inferred in many scenarios; however, the debate about the criticality hypothesis has never ceased. First, it is hard to perceive our brain, which is resilient to many disturbances, can only operate on a critical boundary. In fact, in the brain, the self-organization never precisely reaches the critical state since it is subject to continuous external simulations [25]. This criticality has also been referred as "self-organized quasi-criticality" [26]. Second, it is unclear how individual neurons or synapses infer the entire neural network. In the neural system, collective mechanisms have to be evaluated from an internal perspective, which likely relies on the dynamics of single neurons or synapses and their interactions [2].

Only recently, several solutions have been brought up to resolve this controversy. One possible way is to construct a brain-phase diagram that characterizes the behavior of neuronal complex systems into qualitatively different phases [27]. With the aid of brain-phase diagrams, one can easily tell whether the normal brain states reside on a critical point/boundary or in a phase region. In addition, a phase diagram that has both network information and individual neuron properties may shed light on how local information feeds into the entire network's dynamic behavior. This approach, however, is difficult to implement since it is hard to access the control parameters (i.e., the parameters that need to be tuned for the phase transition) in the experimental settings. For example, it is difficult to carry out experiments where the connectivity of the brain can be varied in *vivo*. Recently, with the advance of neuromorphic chips, with a neural model, it may be now possible to investigate neural dynamics, at least within the approximate model of neurons and their interconnects. Clearly, this kind of chips allows us to conduct experiments freely and to add the understanding of the neural dynamics with the degree of freedom, which could not have been done before. Meanwhile, by understanding the cognitive function of the brain, we may further enable the design of neuromorphic chips to provide yet a powerful new computing paradigm based on the connected networks of spiking neurons [28].

Recently, a mathematical work on the approximation-free supersymmetric theory of stochastic (STS) revealed that, with noise, the general phase diagram of weak-noise stochastic dynamics consist of the three major phases: the thermodynamic equilibrium (T-phase), the noise-induced phase (N-phase, previously known as self-organized criticality), and the ordinary chaos (C-phase) [29]. The phase if present, is of paramount importance for understanding cognitive fractions of a healthy brain as well as its wellness. In order to study the presence of these phases, we used the neuromorphic chip, "Spikey" [30] to emulate different neural dynamic phases and compare them with clinical data of the brain. We demonstrated in this study that under different control parameters of the neuronal firing threshold and external noise level, the behavior of the neural dynamics can be roughly categorized into three phases. To validate our emulation results, we compared our outcomes with clinical recordings from human brain tissues. That we observed similar characteristics showed us that our emulations, within the approximation of the model of neuron, reproduced that of the biological brain. With the flexibility of the artificial "neurons" of the neuromorphic chip, we showed that the early misunderstood "self-organized critical boundary" was in fact a phase induced by noise in the neural dynamics phase diagram based on the power-spectra analysis of many single neuron recordings. Next, we demonstrated another method of constructing the phase diagram using collected neuronal firing activities in the neuronal network. Finally, we envisioned two possible applications of neural dynamics as the result of the study: one is to monitor novel mental wellness and eventually to predict mental abnormality; the other is to use what we learn from the healthy brain in the phase space, or N-phase to construct and perform information processing.

## 2. Settings

The neural dynamics was investigated using a Spikey chip, which used a integrate-and-fire neuron model [28]. First we first considered the controllable available parameters. Table 1 shows the accessible parameters of each individual neuron and synapse in the Spikey chip. In this paper, we used the firing threshold voltage ($V_{th}$) as the controllable variable to tune the dynamical system. It is worth noting that unlike doing experiments on the real brain, it is easy to reconfigure the neuronal network structure (by changing the synapse connection and synapse weights) on the Spikey chip. This gives us more flexibility for future studies.

In this study, we constructed a recurrent network of neurons with sparse and random connections from the Spikey chip. As illustrated in Figure 1(A), a group of 192 neurons in the chip was used. Each neuron was configured to have a fixed number $k = 5$ of presynaptic partners that were randomly drawn from all other hardware neurons. Another set of Eight (8) randomly chosen neurons from the whole group were stimulated by an external input modeled as a uniform-time-distributed signal. The average effect of these external inputs for triggering neuronal firing was 0. In order to have no mean effect under neurons' firing, it was chosen to have half of the external stimuli to be excitatory and half of them inhibitory. The neuron's resting membrane potential was chosen to be -65mV, which is slightly different from the real neuron's resting membrane potential (usually between -75 and 70mV). In our present study, the small difference should not matter as only the difference between the neuronal firing threshold and the resting membrane potential is important. Because of the limit of hardware parameter range, we set the resting membrane potential at -65mV.

## 3. Results

### 3.1: Establish Spikey Validity

- **a. Emulation results from the neuromorphic chip**

In this *Emulation 1* section, we presented three types of neural dynamics. In our emulation setup, the average time interval between two consequent external stimuli was 25ms. By tuning the neuron firing threshold to -57mV, -60mV, and -62mV, respectively, we observed three different neural dynamic behaviors, as illustrated in Figure 1[B~D]. The top row of Figure 1[B~D] show the emulated network activities given by raster plots, while the bottom row includes the membrane potential recordings of a randomly chosen neuron. In Figure 1[B], we observed an intermittent firing behavior from the membrane potential recordings while there was nearly no firing activity in Figure 1[C] and a constant oscillatory firing behavior in Figure 1[D]. Based on these observations, we refer to these three typical dynamic behaviors as Normal firing activity, Coma-like firing activity and Seizure-like firing activity with their corresponding phase states as N-phase (Noise-induced phase), T-phase (Thermal equilibrium phase), and C-phase (Chaotic phase) respectively [30].

Although it seems hard to quantitatively distinguish the differences among those three states from the raster plots at instance, as we will explain later, an order parameter can be established and used to categorize each phase. At this point, we would like to discuss the raster plot qualitatively. In the coma-like state shown in Figure 1[C], there is scarce firing activity from the emulation, and this firing activity does not propagate(no neuronal avalanche happens). This "quiet" behavior is a consequence of the big gap between the firing threshold and the resting potential; thus neurons are seldom fire. In the normal and seizure-like states Figure 1[B, D], however, the firing activities can be observed more often. In Figure 1[B] of the neurons fire intermittently, we can observe this process from the raster plot, while in Figure 1[D], almost every neuron fires continuously.

- **b. Data of the Brain's Recording**

To further validate our findings, we compared our emulation data with recordings from real brain tissues in this section. Neocortical sample sites from pediatric surgery cases were excised for *in vitro* electrophysiological evaluation based on abnormal neuroimaging and electrocorticography (ECoG) assessments. Through electrophysiological and pharmacological methods, the 3 states can be replicated: the coma-like state was observed at the resting membrane potential in the deafferented slice, the normal physiological state can be induced by threshold membrane depolarization, and the seizure activity was induced by blocking $GABA_A$ and $GABA_B$ receptors with bicuculline (10 $\mu$M) and phaclofen (6 $\mu$M), respectively. Note that the small differences between two measurement settings for the brain slice experiments and the Spikey itself caused discrepancies in the results. As listed in Table 2, these differences include measurement sampling rate, recorded signal range, dominant frequency related to different biological time constants, etc. In order to eliminate these discrepancies, we used the FWHM (full-width half-maximum) of the typical firing activity time interval as the unit time to normalize the data. The amplitude of all the data for both sets mapped to 0~1.

To further quantify the three different states, we applied a power spectrum analysis for both the stimulation and the membrane potential recordings of brain tissues. As expected, the power-spectrum behaved quite differently in the frequency domain, which provided us with another powerful tool to analyze the dynamic system. Figure 2 shows the comparison of both clinical and emulation results. We found similar traits in all three states: in the coma-like state (the top row in Figure 2[A]), we did not observe any firing activities at membrane potential recordings in both clinical and emulation settings. Their corresponding power-spectra showed a sharp drop at low frequencies and the typical normalized power-spectra (the top row in Figure 2[B]) in the frequency domain are below $10^{-3}$. In the normal state (the middle row in Figure 2[A]) we observed intermittent firing activities in both measurements. Their power-spectra (the middle row in Figure 2[B]) showed a $1/f^\alpha$ noise-like behavior. In the seizure-like state (the bottom row in Figure 2[A]), we observed an oscillatory behavior and found that the power-spectra (the bottom row in Figure 2[B]) show $1/f^\alpha$ noise-like behavior superimposed by some frequency at low peaks.

These results showed that the stimulation provides a reasonable description of the activity of the real brain at least, within the approximation model used to first-order approximation. With the neuromorphic chip, we were able to further manipulate the dynamics in the N-phase by varying the parameters, indicating the N-phase has a width, related to the noise as to be discussed next. For more experimental details, please refer to Materials and Methods.

## 3.2: Constructing brain phase diagram

In this section, the dynamics phase diagram is constructed using the Spikey emulation by controlling the firing threshold and noise level. Owing to the high configurability of the chip, we can explore a relatively large controllable parameter space. Recently, our early work on the approximation-free supersymmetric theory of stochastic (STS) [28] revealed that, with noise, the "early conceived" "self-organized criticality" is misunderstood and predicted that the general phase diagram of weak-noise stochastic dynamics on separate phase of the brain neuronal networks should consist of the three major phases: the thermodynamic equilibrium (T-phase), the noise-induced phase (N-phase, previously known as self-organized criticality), and the ordinary chaos (C-phase).[29]

To validate this picture, we performed emulation studies on the Spikey. On the basis of the settings as described in *Emulation Settings*, we also included the 'Noise Level' as a controllable parameter. Specifically, the induced noise intensity was applied in the time interval between two subsequent noise stimuli and the noise stimuli were uniformly distributed during the whole emulation time (1000 ms).

The resulting phrase diagrams given on the plane of the firing threshold potential $V_{th}$ and the noise intensity is illustrated in Figure 3, which were constructed through a power-spectra analysis of the membrane potential recordings. The classifications of the phases are similar to that described in Section 3.1based on the power-spectra analysis applied to the membrane potential recordings, was shown. When there was no noise presented in the system, there were only two phases: the T-phase, with no conspicuous dynamics, and the C-phase featured by oscillatory chaotic activities. The inset figure, zooming the sharp transition between these two phases, shows that the C-phase featured by power-spectra with equidistant peaks superimposed on the $1/f^\alpha$ occurs at $V_{th}$ =-60.6mV, while there was no spiking activity at $V_{th}$ =-60.5mV. The transition was abrupt due to the fact that 0.1 mV is the smallest resolution for the Spikey chip; we could not get a sharper phase transition. As the noise intensities increased, another distinct noise-induced phase (N-phase) as discussed earlier shows up, The border of the N-phase was highly dependent of the noise intensity where a higher noise intensity results in a wider N-phase (in the sense of threshold range). From this simple emulation study it is clear N-phase (which exists in normal healthy brain), should reside in a certain full-dimensional phase rather than at a critical boundary.

## 4. Discussion

In this section, we will show an alternative way to construct the phase diagram and discuss how this neural dynamics phase diagram may provide perspectives in metal disorder diagnosis and treatment, information processing method, etc.

### 4.1 An alternative way to construct the phase diagram

In order to take the entire neural network activity into account and give a more completed picture, we used an alternative way of analysis to reconstruct the brain phase diagram. Previous studies [31, 32, 33] have shown that it is advantageous to quantify the neural dynamics by defining an order parameter $<C_{sync}>$ as the average over the correlations between pairs of neurons $i,j$, as

$$C_{i,j}(\tau) = \frac{1}{\tau} \int_{t_0}^{t_0+\tau} \delta_i(t)\delta_j(t)\, dt \qquad [1]$$

where $\delta_i(t)$ is 1 if neuron $i$ is spiked at time $t$, and 0 otherwise. We can extract this parameter from the raster plots shown in Figure 1. [B~D].

The 3D-plot of the order parameter $<C_{sync}>$ on the threshold-noise level plane is shown in Figure 4. For all noise levels, $<C_{sync}>$ increases as the threshold decreases. At the threshold around $V_{th}$=-60.5mV, we see that $<C_{sync}>$ becomes pronounced (> 10$^{-3}$), i.e. a phase transition to the C-phase (Red & pink colors), at all noise levels. This is exactly the same results as we observed for the phase transition based on the power-spectra analysis. When the noise intensity increased, we observed that a 'plateau phase', namely the 'N-phase' (green), shows up. The magnitude of $<C_{sync}>$ of the plateau in the middle at all noise levels was around 10$^{-4}$, which was depicted by green color; while the T-phase (indicated by the gray color) without conspicuous firing activities usually had a correlation parameter below $\sim 10^{-6}$, two orders of magnitude smaller. Table 3 summarizes the typical values for $<C_{sync}>$ in the different phases.

Given the similarity of phase-diagram derived from both the power-spectra method and the correlation parameter method, it is clear that the local neuron activity can reflect global dynamical states. In another words, when the single neuron fires continuously, it 'feels' a higher activities probability of the entire system which resides in the C-phase. Using this argument, the single neuron is tuned by its network environment until the whole network settles down into a phase. This fact, to some extent, can explain how the individual neuron or synapse and the local observation of such can be useful to infer the whole network properties. Details of this part is given in refer to *Materials and Methods*.

## 4.2 Potential applications of neural dynamics

One natural question that may arise is why exactly the nature has chosen for the normal brain state to reside in the N-phase. We do not have a satisfactory answer for this question yet. There is, however, a line of reasoning that seems to answer this question partially. It was recently argued that natural dynamical systems in the N-phase may be the most efficient natural optimizers [34]. If this point of view is adopted, then it could be speculated that, at least, one of the functionalities of neural dynamics is to optimize the performance of information processing. The main significance of this study compared to for example the use of neural networks in deep learning, is that the dynamics may be used to create a different neuromorphic computing paradigms. Indeed, to our knowledge, this is the first time we propose the neuromorphic computing from a dynamical system point of view.

More effort is needed in this direction in order to better understanding as of how neural dynamics can optimize complex tasks such as to solve complex problem, decision making, and eventually complex cognitive and intelligent functionalities. If the mechanism can be illustrated more clearly, neuromorphic computing will gain a significant advantage compared to traditional software solutions, since it has an inherent stochastic (noise) property in the physical system that is essential for the N-phase to exist.

Another potential application is towards a better way to monitor and treat mental disorders. As we mentioned previously, the three different phases can be dubbed as the "coma-like", the "normal" and the "seizure-like" phases, respectively. With these specific characteristics of neural dynamics in mind, doctors and patients will gain more information about when the disease onsets and what is the better way to cure and prevent the disease from happening. Of course, the brain is much more complicated than a simplified stimulation. Clearly, much more complicated dynamics phase diagram and a more sophisticated phase-diagram with finer structures and subphases can be developed in the future. Furthermore, the synapses are important. In particular, the interactions of neural dynamics and spikes are equally important to further understand the functionality of the brain. This work represents an important beginning of the collective neural dynamics.

## 5. Conclusion

In this paper, we have discussed neural dynamics of clinical recordings and emulations on a neuromorphic chip. We first presented the three typical phase states, namely the T-phase state, the N-phase state, and the C-phase state, featured by no conspicuous dynamics, intermittent firing dynamics and constant oscillating activities, respectively. We utilized the configurability in the Spikey chip to construct a brain phase diagram by varying dynamic parameters, firing threshold and noise level plane. This phase diagram is supported by the recently enunciated STS theory. An alternative method was also used to construct a similar phase diagram using the neuronal firing activities in the entire network. We also envision two possible applications using neural dynamics: one is to pave a new way for mental disorder diagnostics and treatments; the other is to use N-phase dynamics to perform information processing. A more comprehensive, detailed of dynamics phase diagram can be presumably constructed in the future once we know the dynamics variable of the brain. The details of the brain dynamic phase can further accelerate our understanding of cognitive functionalities as well as wellness.

## Materials and Methods

### Brain Slice Experiments

In the real brain slice experiments, neocortical sample sites were excised for *in vitro* electrophysiological evaluation based on abnormal neuroimaging and electrocorticography (ECoG) assessments. Sample sites ($2\,\text{cm}^3$) were removed microsurgically and directly placed in an ice-cold artificial cerebrospinal fluid (ACSF) containing (in mm); NaCl 130, $NaHCO_3$ 26, KCl 3, $MgCl_2$ 5, $NaH_2PO_4$ 1.25, $CaCl_2$ 1.0, glucose 10 (pH 7.2-7.4). Within 5-10 min, slices (350 $\mu$m) were cut (Microslicer, DSK Model 1500E or Leica VT1000S) and placed in ACSF for at least 1h (in this solution $CaCl_2$ was increased to 2 mm and $MgCl_2$ was decreased to 2mM). Slices were constantly oxygenated with 95% $O_2$-5% $CO_2$ (pH 7.2-7.4, osmolality 290-300 mOsm, at room temperature). Methodological details for cell visualization and identification have been published elsewhere [29]. Patch electrodes (3-6 MΩ) were filled with (in mM) Cs-methanesulfonate 125, NaCl 4, KCl 3, $MgCl_2$ 1, MgATP 5, ethylene glycol-bis (β-aminoethyl ether)-N,N,N',N'-tetraacetic acid (EGTA) 9, HEPES 8, GTP 1, phosphocreatine 10 and leupeptine 0.1 (pH 7.25-7.3, osmolality 280-290 mOsm) for voltage clamp recordings or K-gluconate 140, HEPES 10 $MgCl_2$ 2, $CaCl_2$ 0.1, EGTA 1.1, and 2 $K_2ATP$ (pH 7.25-7.3, osmolality 280-290 mOsm) for current clamp recordings. The access resistance ranged from 8-20 MΩ. Liquid junction potentials (~6 mV) were not corrected. Cells were initially held at -70 mV in the voltage clamp mode. At this holding potential, the passive membrane properties in the slices were determined by applying a depolarizing step voltage command (10 mV) and using the membrane test function integrated in the pClamp (version 8) software (Axon Instruments, Foster City, CA). This function reports membrane capacitance (in pF), input resistance (in MΩ) and time constant (in ms). As paroxysmal discharges rarely occur spontaneously in slices from the cortical tissue samples, this activity was induced by a combination of the $GABA_A$ receptor antagonist bicuculline (20 μM) and the K-channel blocker 4-aminopyridine (100 μM). Ictal activity was recorded after blockade of the $GABA_B$ receptors with the antagonist phaclofen (6 μM).

### Emulation and Medical data comparison

Having noted that the difference between the ranges of spiking signal in the two systems was too large for effective comparison, we mapped the membrane potential of both medical and emulated data to a scale of 0 to 1. The process has three steps: 1) the graph of membrane potential vs. time was shifted vertically by subtracting the minimal membrane potential from the original data of all three phases; 2) The data cursor tool was used to determine the amplitude of the subtracted membrane potential of the seizure phase; 3) The membrane potential signal for the three phases was re-scaled to 0~1.

Due to the intrinsic spiking frequency difference, the need to normalize the data arises prior to the comparison between the medical and emulated data. In order to compare the shape of power spectra for the medical data and emulated data, the time series membrane potential recordings, whose signal range have been mapped from 0 to 1, were normalized with unit spike width. The unit spike width is $\Delta t_1$ for medical data and $\Delta t_2$ for emulated data. As shown in Figure 5, the unit spike width was defined as the full width at half maximum of a spike on the graph of membrane potential against time. The time coordinates of the end-points were determined using the data cursor tool in Matlab.

Standard power-spectra analysis was then applied to the normalized time-series membrane potential data.

**Correlation parameter extraction**

In order to deliver a more robust and convincing discussion on the whole picture, we used an alternative approach to construct the brain phase diagram. Previous studies [31, 32, 33] have shown that it is advantageous to quantify the brain dynamics by defining an order parameter $< C_{sync}(V_{th}) >$ as the average over the correlations between pairs of neurons $i,j$ shown in Eq[1]. For computational purpose, we discrete the equation as follows:

$$C_{i,j}(\tau) = \frac{1}{\tau} \sum_{t=t_0}^{t_0+\tau} \sigma_i(t)\sigma_j(t) \qquad [2]$$

where $\sigma_i(t)$ is 1 if the neuron $i$ spiked at time $t$, and zero otherwise. This quantity was evaluated over a time window of $\tau = 5\ ms$ in our experiments, which was approximately as the typical stimulation time period of the chip.

Since in reality it is difficult to observe the spiking of neurons at a fixed discrete time value, we divided the time window $\tau$ into 5 time bins. The parameter was extracted from the raster plot as shown in Figure 6. Each point $(i, t)$ in the raster plot represented the spiking activity of neuron $i$ at time bin $t$. To extract this parameter, a matrix representation of this raster plot was first constructed using Matlab. In our experiment, there are 192 neurons and each neuron was observed over a time period of 1000 ms. Hence, we constructed a 192 × 1000 zero-one matrix $\mathbf{\Sigma}$ to represent each neuron's spiking activity as shown in the raster plot. Each row corresponded to a neuron whereas each column corresponded to a time bin. If the neuron $i$ spikes at time bin $t$, we denote it as $\mathbf{\Sigma}_{it} = 1$.

The parameter $C_{i,j}$ for each pair of neurons $i,j$ can then be extracted from matrix $\mathbf{\Sigma}$ in three steps: 1) perform element-wise multiplication on the row vectors $\mathbf{\Sigma}_i$ and $\mathbf{\Sigma}_j$ to obtain a resultant vector, denoted as $m_{ij}$, 2) sum up the entries in $m_{ij}$ and divide by a constant $\tau$ as indicated in the equation, 3) divide the results obtained in previous step by 200, given there are 200 windows, to obtain the average correlations at certain conditions (threshold noise level). Lastly, given a total of 192 neurons, there were 18,336 pairwise correlations $C_{i,j}$, $1 \leq i \neq j \leq 192$, at each threshold level. The average over all the correlations $C_{i,j}$ was the value of $<C_{sync}(V_{th})>$ at threshold level $V_{th}$ as shown in Figure 4.

**ACKNOWLEDGEMENTS.** K.L.W. would like to acknowledge the support of the endowed Raytheon professorship. The neuromorphic hardware and software is partially supported by EU Grant 269921 (BrainScaleS), and EU Grant 604102 (Human Brain Project, HBP). We would like to thank Thomas Pfeil for his support of the hardware system.

Table 1: List of the accessible parameters for the individual neurons and synapses on the Spikey chip. Spikey uses a simple integrate-and-fire model of a neuron. Redrawn from [28]. Each parameter is shown with the corresponding model parameter names and their corresponding description. Electronic parameters that have no direct translation to the model parameters are denoted *NA*.

| *Scope* | *Name* | *Type Description* |
|---|---|---|
| Neuron circuits | $\tau_{refrac}$ | *Bias current controlling neuron refractory time* |
|  | $V_{th}$ | *Firing threshold voltage* |
|  | $V_{reset}$ | *Reset potential* |
| Synapse and line drivers | NA | *Two digital configuration bits selecting input of line driver* |
|  | NA | *Two digital configuration bits setting line excitatory or inhibitory* |
|  | $t_{rise}, t_{fall}$ | *Two bias currents for rising and falling slew rate of presynaptic voltage ramp* |
|  | $I_{rise}^{max}$ | *Bias current controlling maximum voltage of presynaptic voltage ramp* |
|  | w | *4-bit weight of each individual synapse* |

Table 2: Comparison of the parameters in the two measurement systems: Emulation on Spikey and experiment on human brain slice.

|  | **Emultioan on Spikey** | **Experiment on brain slice** |
|---|---|---|
| **Measurement method** | — | Current clamp measuring |
| **Measured subject** | 1 of 192 neurons in Spikey | 1 neuron in human brain |

|  | neural network | slice |
|---|---|---|
| **Measurement sampling rate** | ~100 µs | 200µs |
| **Observable highest frequency** | ~5 kHz | 2.5 kHz |
| **Recorded signal range** | -75mV ~ -60mV | -80mV ~ 60mV |
| **Dominant frequency** | 0.1kHz | 0.02kHz |

Table 3: Typical $<C_{sync}>$ under different phases.

| **Phase** | **Typical** $<C_{sync}>$ |
|---|---|
| T-phase | $< 10^{-6}$ |
| N-phase | $\sim 10^{-4}$ |
| C-phase | $> 10^{-3}$ |

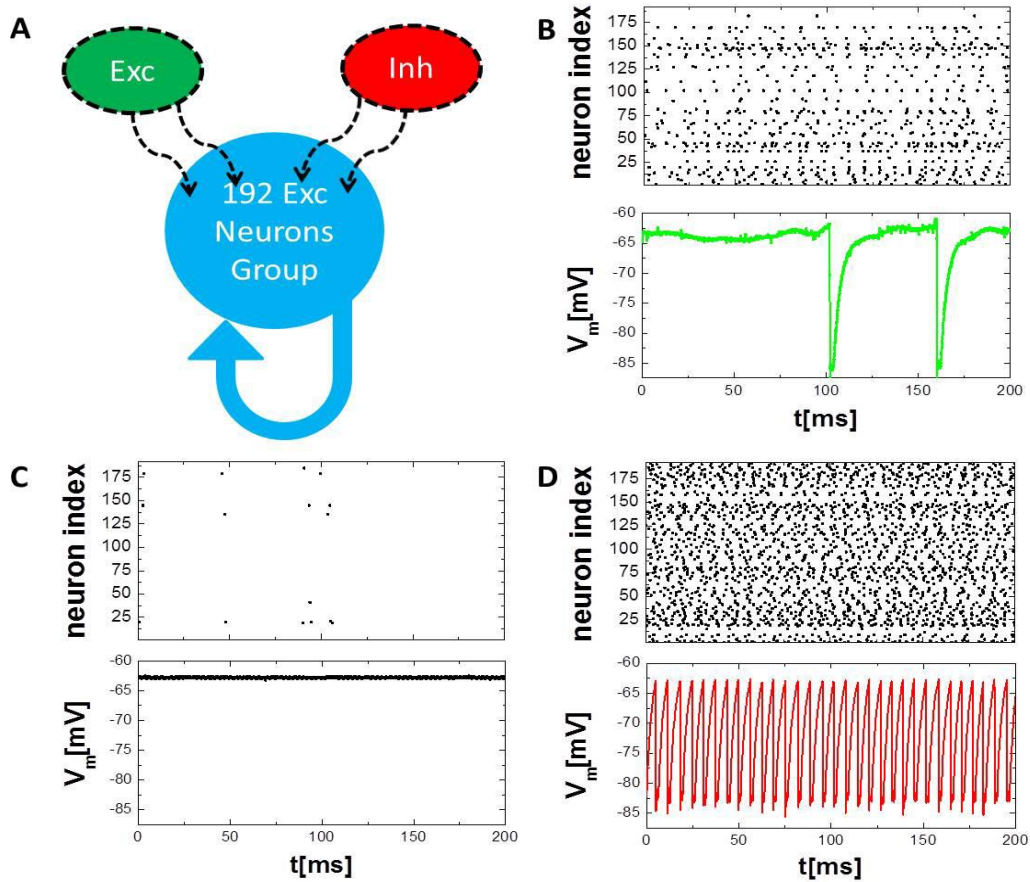

Fig. 1: [A] Illustration of emulation settings on the Spikey chip. A group of 192 neurons was used to form a recurrent neural network. In order to have no mean effect under neurons' firing, it was chosen to have half of the external inputs to be excitatory and half of them inhibitory; [B~D] Three typical neural dynamical behaviors in the neural network; top: the raster plots of neuronal activities in the neural network; bottom: the membrane potential recordings of one randomly chosen neuron. [B] Normal state / N-phase behavior shows an intermittent firing activity with relatively low activity correlation (mainly the precedence relationship); [C] none or only a few firing activities present in coma state / T-phase with extremely low activity correlation; [D] A constant oscillating firing behavior present in the seizure state / C-phase with a high activity correlation.

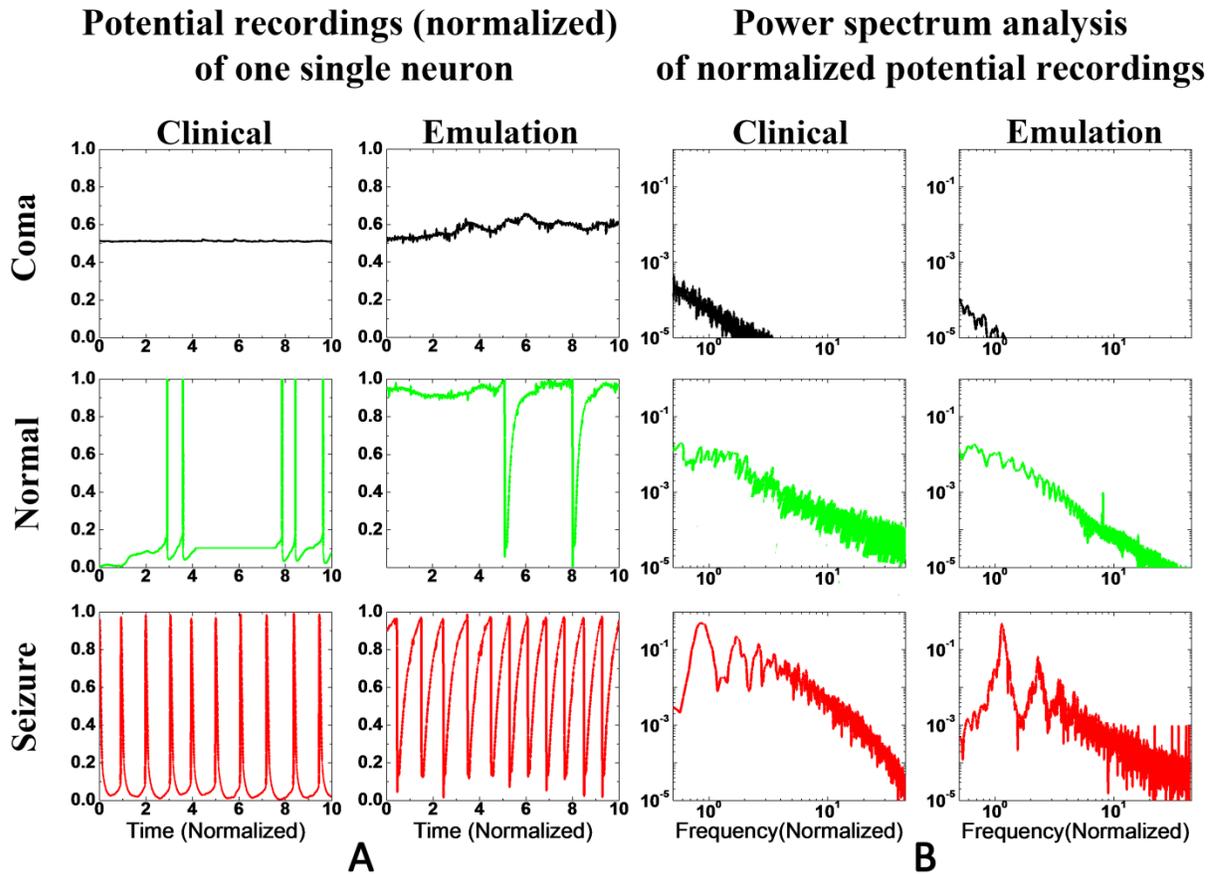

Fig. 2: Comparison between human brain slice data and emulation data in the coma-like (top line), the normal (middle line) and the seizure (bottom line) states. In the coma-like state, we observed no firing activity at both membrane potential recordings (Top left A). Their correspondent power-spectrum shows a sharp drop at low frequencies and the typical normalized power-spectrum in the frequency domain is below $10^{-3}$ (Top right B). In the normal state, we observed the intermittent ring activities in both systems (Mid left A). Their power-spectra shows a $1/f^{\alpha}$ (a linear line on log-log plot) behavior (Mid right B). In the seizure-like state, we observed the oscillatory behavior from both membrane potential

recordings (Bottom left A) and found the power-spectra are both the $1/f^{\alpha}$ noise-like superimposed by peaks (Bottom right B).

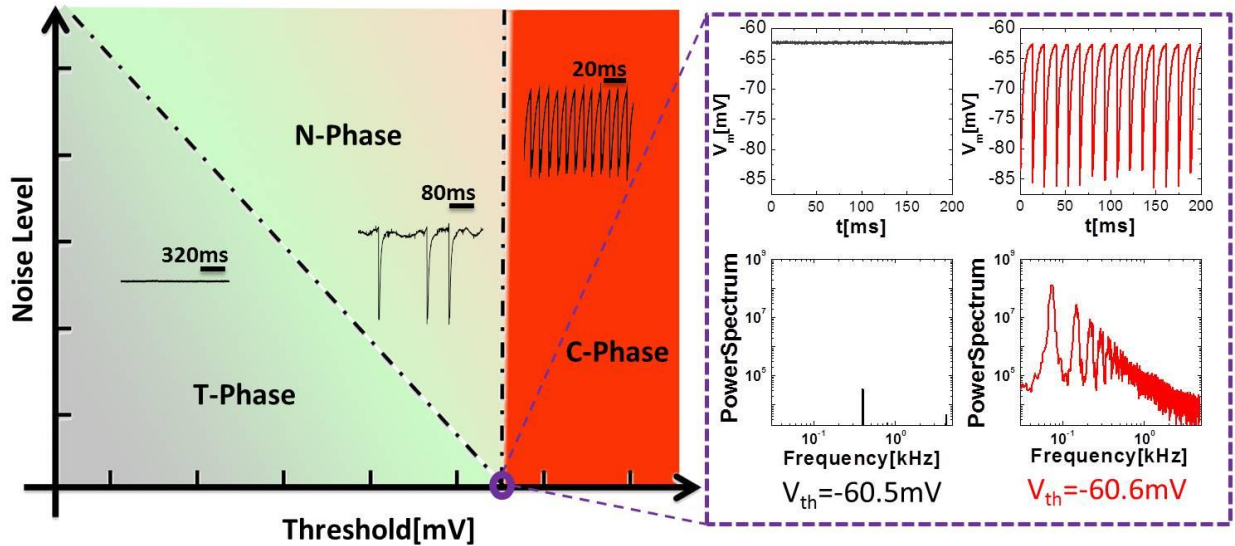

Fig. 3: Phase diagram of neural dynamics constructed based on power-spectra analysis. The results show that in the deterministic limit (no noise present), the N-phase collapses onto a sharp transition between the T-phase and C-phase (vertical dashed line at around -60.5 mV), as predicted by the STS. The inset figure zooms in on the sharp transition between these two phases, from $V_{th}$=-60.6mV to $V_{th}$=-60.5mV. When noise is present in the system, the noise-induced phase (N-phase) shows up. The noise is an essential parameter in this picture. The higher the noise intensity is, the wider the N-phase (in the sense of threshold range) become.

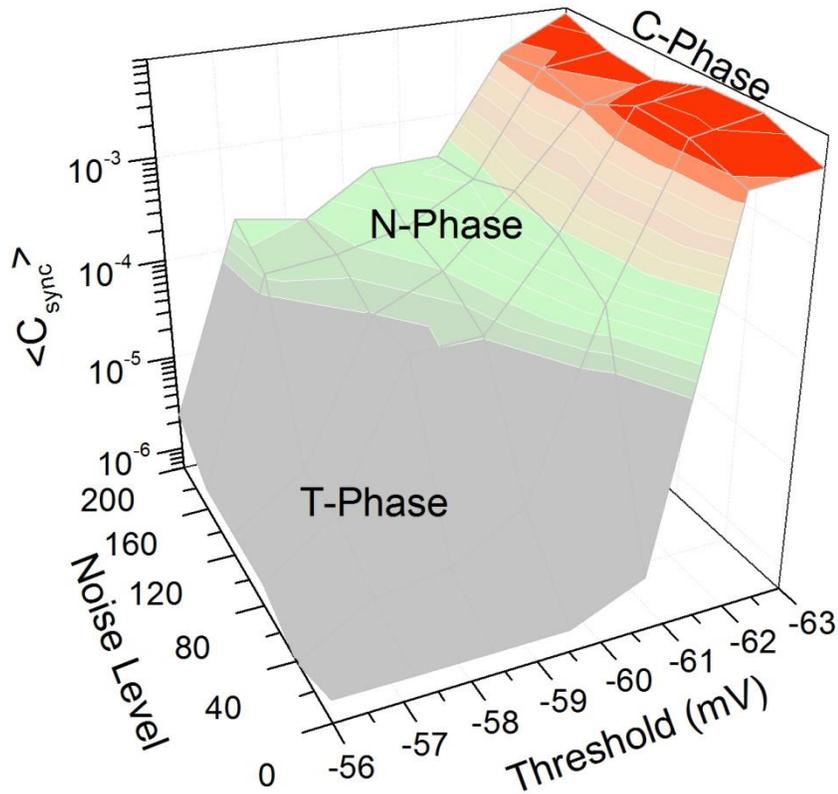

Fig. 4: 3D-plot of $<C_{sync}>$ in the threshold-noise level plane, an alternative way to construct the phase diagram. Based on the correlation order parameter $<C_{sync}>$, a similar brain phase diagram was constructed as that in Figure 3. For all noise levels, $<C_{sync}>$ increases as the threshold decreases. At a threshold around $V_{th}=-60.5mV$, we see that $<C_{sync}>$ becomes pronounced ($>10^{-3}$), i.e. a phase transition to the C-phase, at all noise levels. The results are exactly the same as the phase transition constructed from the power-spectra analysis. When the noise intensity increases, we observed that a 'plateau phase', namely the 'N-phase', shows up. The magnitude of $<C_{sync}>$ at the plateau at all noise levels is around $10^{-4}$, which is depicted by the green color in the picture; while the T-phase (indicated by the gray color) without conspicuous firing activities usually has a correlation parameter below $10^{-6}$, two orders of magnitude smaller.

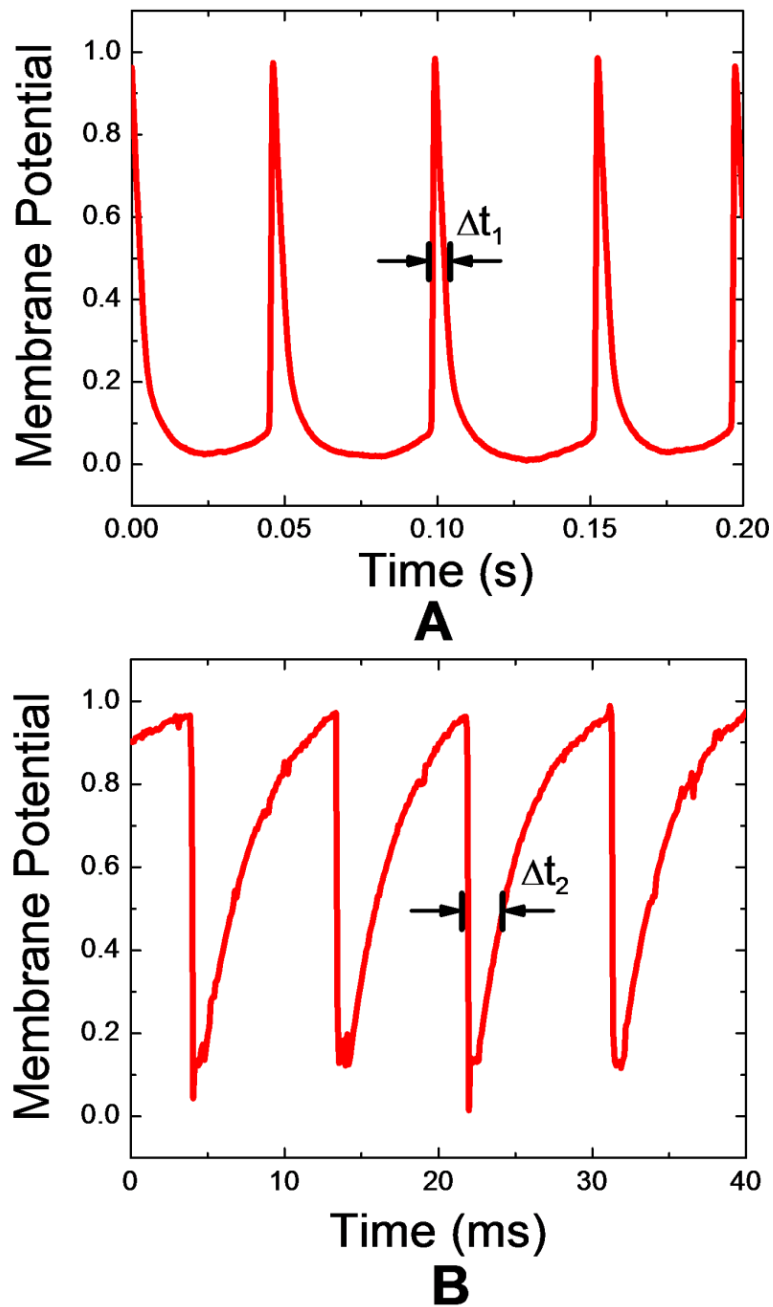

Fig. 5: The full width at half maximum is used to define the unit spike time for two different sets of data. [A] Determine Δt$_1$ for medical data; [B] Determine Δt$_2$ for emulated data. The different sign of the membrane potential change during firing comes from the difference of the neuron model that uses in Spikey chip and the real neuron behavior. In Spikey,

integrated and fire model is used for each neuron, thus after each spike, the membrane potential of the neuron will be reset to its rest potential. Thus the sign of [B] was opposite to [A].

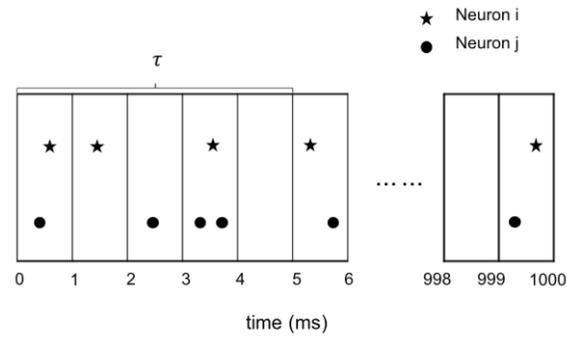

Fig. 6: The raster plot of neurons i,j over 1000 ms. Each point represents the occurrence of neuron spiking. The row vectors corresponding to the raster plot of neuron $i$ and $j$ are [1,1,0,1,0,1,…,0,1] and [ 1,0,1,1,0,1,…,0,1] respectively.